\documentclass{pasj00}
\draft

\begin{document}
\SetRunningHead{Y. Shioya et al.}{NB921 dropouts in SDF}
\Received{2005/mm/dd}
\Accepted{yyyy/mm/dd}

\title{
A survey of NB921 dropouts in the Subaru Deep Field
}

\author{Yasuhiro \textsc{Shioya},\altaffilmark{1} 
Yoshiaki \textsc{Taniguchi},\altaffilmark{1} 
Masaru \textsc{Ajiki}, \altaffilmark{1} 
Tohru \textsc{Nagao}, \altaffilmark{2,3} \\
Takashi \textsc{Murayama}, \altaffilmark{1} 
Shunji S. \textsc{Sasaki},\altaffilmark{1} 
Ryoko \textsc{Sumiya},\altaffilmark{1}
Yuichiro \textsc{Hatakeyama},\altaffilmark{1} \\
Taichi \textsc{Morioka},\altaffilmark{1} 
Asuka \textsc{Yokouchi},\altaffilmark{1}
Osamu \textsc{Koizumi},\altaffilmark{1} 
Mari I. \textsc{Takahashi},\altaffilmark{1} \\ 
\& 
Nobunari \textsc{Kashikawa} \altaffilmark{2}

}
\email{shioya@astr.tohoku.ac.jp}

\altaffiltext{1}{Astronomical Institute, Graduate School of Science, Tohoku University, \\
                 Aramaki, Aoba, Sendai 980-8578}
\altaffiltext{2}{National Astronomical Observatory of Japan,\\
                 2-21-1 Osawa, Mitaka, Tokyo 181-8588}
\altaffiltext{3}{INAF --- Osservatorio Astrofisico di Arcetri,\\
                 Largo Enrico Fermi 5, 50125 Firenze, Italy}



\maketitle

\begin{abstract}
In order to search for high-redshift galaxies beyond $z = 6.6$ 
in the Subaru Deep Field, we have investigated NB921-dropout
galaxies where NB921 is the narrowband filter centered at 
919.6 nm with FWHM of 13.2 nm for the Suprime-Cam on the Subaru
Telescope. There are no secure NB921-dropout candidates brighter 
than $z^\prime = 25.5$. Based on this result, we discuss the UV 
luminosity function of star-forming galaxies at $z > 6.6$. 

\end{abstract}

\section{Introduction}

In the last decade, surveys for high-redshift galaxies have been 
very successful thanks to the great observational capabilities of
8-10m class telescopes. Both the Lyman break method and the
Lyman $\alpha$ emitter search technique have provided us a large number of
high-redshift galaxies at $z > 3$ (e.g., Steidel et al. 1996, 1999; 
Ouchi et al. 2004a, 2004b; Hu et al. 2002; Ajiki et al. 2003; 
Kodaira et al. 2003; Taniguchi et al. 2005). 
In particular, these surveys  tell us the evolution of the star formation
rate density up to $z \sim 6$ (e.g., Taniguchi et al. 2005;
see also for review, Taniguchi et al. 2003; Spinrad 2004) and 
the presence of large scale structure in the very early universe 
(e.g., Ouchi et al. 2005 and references therein).

Now, the frontier of high-redshift galaxy surveys have been shifted to $z > 6$.
It becomes important to clarify the star formation rate density and 
the clustering properties at $z > 6$. 
Spectroscopic observations of the highest known quasars suggest 
that the reionization of the intergalactic medium (IGM) could have just completed 
at $z \ge 6$ (Djorgovski et al. 2001; Becker et al. 2001; see for a review 
Loeb \& Barkana 2001). 
On the other hand, a large amplitude signal in the temperature-polarization 
maps of the cosmic microwave background by The Wilkinson Microwave Anisotropy Probe (WMAP) 
is interpreted as that the universe became reionized at $z \sim 17$ 
(Spergel et al. 2003; Kogut et al. 2003). This also reinforces the importance of 
investigation of galaxies beyond $z \sim 6$ from various observational points
of view.

Recently, Bouwens et al. (2004b, 2005) tried to search galaxies at $z \sim$ 7--8 
as $z_{850}$-dropouts in the Hubble Ultra Deep Field (HUDF) and  
those at $z \sim 10$ as $J_{110}$-dropouts in the fields with deep NICMOS image. 
They found four and less than three candidates, respectively, in their surveys.
These pioneering works may lead to a conclusion that 
the star formation rate density is declining toward higher redshift
beyond $z \sim 6$. 
Although the HUDF surveys are unprecedentedly deep, their survey field is not wide enough
to obtain a general conclusion because the field of view of the NICMOS is very small. 
We therefore tried to search galaxies at $z > 6.6$ 
using the deep optical imaging data set for the Subaru Deep Field
(SDF: Kashikawa et al. 2004). Applying the NB921-dropout method, we
investigate how many such very high-$z$ galaxies are present in this field.
We describe the NB921 dropout method in section 2. 
In section 3, we then apply the method to the SDF data 
We discuss the star formation rate density at $z > 6.6$ in section 4. 

\section{NB921 dropouts}

The dropout method is a powerful tool to find high redshift galaxies. 
The basic idea of the NB921-dropout method used in this study is almost the same as
that used in previous dropout galaxy surveys
(e.g., Steidel et al. 1996, 1999; Stanway et al. 2004). 
Since the UV continuum shortward  Ly$\alpha$ of high-$z$ galaxy is absorbed 
by the intergalactic neutral hydrogen gas clouds, 
we can use this property to select high-$z$ galaxies. 
We call galaxies with $z > 6.6$ as (genuine) NB921-dropout galaxies and 
galaxies with $NB921-z^\prime > 0$ simply NB921-depression galaxies. 
Note that a similar narrowband dropout method has been already applied to 
the SDF data; the NB816 dropout method to select galaxies with $5.7 < z < 6.5$ 
(Shioya et al. 2005b; see also Shioya et al. 2005a).

In fig.1, we show the transmission curves of the two filters of $z^\prime$ and NB921. 
We also show SEDs of an LBG at $z = 6.1$, 6.57, and 6.8. 
The stellar continuum of an LBG is produced by GALAXEV (Bruzual \& Charlot 2003). 
Here we adopt $\tau$ model, i.e., $SFR(t) \propto \exp(-t/\tau)$ with $\tau=1$ Gyr and 
age of $t=1$ Gyr. 
We then add the Lyman $\alpha$ emission line with $EW_0({\rm Ly}\alpha)=65$ \AA 
to the above stellar continuum. We also use the cosmic transmission by Madau (1996)
to evaluate the absorption by intergalactic medium. 
Note, however, that we ignore the extinction by dust in this figure. 

Since the bandpass of the NB921 filter is completely included in that of the 
$z^\prime$ filter, the relation between $NB921-z^\prime$ and $z$ is slightly complex. 
For example, as we mentioned in Taniguchi et al. (2005), 
if the UV continuum depression shortward of Lyman $\alpha$ (121.6 nm) 
due to the absorption of the intergalactic medium is redshifted 
not into the NB921 window but into  the $z^\prime$ window, corresponding to a
case that galaxies are located at $5.9 < z < 6.5$, 
the $NB921-z^\prime$ color becomes to be bluer (i.e., NB921-excess). 
However, as we mentioned in Nagao et al. (2004), 
even if this is the case, when a galaxy has a strong Ly$\alpha$ emission line, 
the $NB921-z^\prime$ color may become redder (i.e., NB921-depression)
although the degree of this depression depends on the strength of Ly$\alpha$ emission. 

Adopting both the cosmic transmission derived by Madau et al. (1996) 
and the extinction curve derived by Calzetti et al. (2000), 
we calculate the $NB921-z^\prime$ color as a function of $z$ 
for the case of $EW_0({\rm Ly} \alpha) =$ 0 (stellar continuum only), 65, 130, 
and 260 \AA, respectively. The results are shown in figure 2. 
It is clearly shown that Lyman break galaxies with strong Ly$\alpha$ emission 
are identified as NB921-depression objects even if their redshifts are smaller than 6.6.
Here we note that $i^\prime$-dropout galaxies with NB921-depression found by Nagao et al.
(2004, 2005) are Lyman$\alpha$ emitters at $z \sim 6$  and their $EW_0({\rm Ly}\alpha)$
ranges from 110 \AA to 280 \AA. We therefore adopt a color criterion of 
$NB921-z^\prime > 1.5$ for unambiguous NB921-dropout galaxies. 

\section{Application to the Subaru Deep Field Data}

We apply the NB921-dropout method to the SDF data (Kashikawa et al. 2004). 
Its photometric catalog (Version 1) is in public\footnote{http://soaps.naoj.org/sdf/}. 
In the following analysis, we use $2^{\prime \prime} \phi$ aperture 
magnitudes and photometric errors in a $2^{\prime \prime} \phi$ aperture 
magnitude in the $z^\prime$-band selected catalog. 
We also calculate errors of colors using the above photometric errors.

We adopt the following selection criteria 
for NB921-dropout galaxies,
\begin{equation}
NB921 - z^\prime > 1.5
\end{equation}
and 
\begin{equation}
z^\prime < 25.48, 
\end{equation}
since we calculate the lower limit of $NB921-z^\prime$ using 
the $2 \sigma$ limiting magnitude for galaxies fainter than the above $z^\prime$
limiting magnitude;  the $2 \sigma$ limiting magnitude of NB921 is 26.98 mag. 
Figure 3 shows a diagram between $NB921-z^\prime$ and $z^\prime$. 
It is found that 14 galaxies satisfy the above criteria (open crosses in figure 3). 

In order to reduce any contaminations from foreground objects 
that are free from absorption by the intergalactic neutral hydrogen, 
we also adopt the following additional criteria
\begin{eqnarray}
B & > & 28.45 \; (3 \sigma), \\
{\rm and} \; \; V & > & 27.74 \; (3 \sigma).
\end{eqnarray}
By using the above three criteria, we located no NB921-dropout object. 
We therefore conclude that there is no galaxy at $z > 6.6$ with $z^\prime < 25.48$
in the SDF. 
All the objects with $NB921-z^\prime > 1.5$ are detected both in $B$ and in $V$ bands 
and considered to be low redshift galaxies. 

Galaxies at $z > 6.6$ could be selected as a subsample of $i^\prime$-dropout galaxies. 
We now investigate the $NB921-z^\prime$ color of $z^\prime$-dropouts. 
We select $i^\prime$-dropout galaxies adopting the following criteria: 
$i^\prime - z^\prime > 1.5$, 
$z^\prime < 26.07 \; (5 \sigma)$, 
$B > 28.45 \; (3 \sigma)$, 
$V > 27.74 \; (3 \sigma)$, and
$R > 27.80 \; (3 \sigma)$. 
The number of $i^\prime$-dropout candidates is 49. 
We plot them on a diagram between $NB921-z^\prime$ and $z^\prime$ 
(figure 3) and on another diagram between $i^\prime - z^\prime$ and 
$NB921-z^\prime$ (figure 5). 
All the $i^\prime$-dropouts except three are detected in NB921. 
For the three galaxies undetected in NB921 (open circles in figure 3), 
we cannot judge if they are genuine NB921-dropout galaxies
unless future spectroscopic confirmation is available. 
However, we conclude that there are  three possible genuine NB921-dropout
candidates in the SDF  if we lower the $z^\prime$ limiting magnitude
down to 26.07. 

Finally, we comment on another three galaxies whose $NB921-z^\prime$ color 
is larger than 0.9 together with above $3 \sigma (NB921-z^\prime)$
(open squares in figure 3). As we discussed in section 2, 
strong Lyman $\alpha$ emitters at $6.0 < z < 6.5$ have a color of 
$NB921-z^\prime \simeq 1$. 
We, therefore, do not select them as NB921-dropout galaxies in this paper, 
although they may be genuine NB921-dropout galaxies at $z \sim 6.6$. 
It is noted that all these galaxies are detected in NB921 and some of 
$z^\prime$-excess objects  are spectroscopically confirmed as Lyman $\alpha$ 
emitters at $z < 6.6$ (Nagao et al. 2004, 2005). 

\section{DISCUSSION}

\subsection{UV Luminosity Function of Galaxies at $z > 6.6$}

We discuss how our new result gives a constraint on the UV luminosity function
of star-forming galaxies at $z > 6.6$. In the present study, we find no
genuine NB921-dropout galaxy in the SDF from a sample of galaxies with
$z^\prime < 25.48$ and three possible candidates from a sample of
$z^\prime < 26.07$.
The observed $z^\prime$-magnitude of a galaxy depends 
not only on the UV luminosity but also on both the redshift of the galaxy and 
the strength of the Ly$\alpha$ emission line. 
Therefore, it seems convenient to show how our upper limits are
consistent with predictions from previous luminosity functions
for high-$z$ galaxies. 

In figure 6, we show  expected cumulative numbers of galaxies at $z > 6.6$
if the UV luminosity function of galaxies at $z > 6.6$ is 
the same as that at $z \sim 4.0$ (Ouchi et al. 2004a) or at $z \sim 6$ 
(Bouwens et al. 2004a) for the following two cases; no Lyman$\alpha$ emission 
or Lyman$\alpha$ emission whose rest-frame equivalent width of 65 \AA.
Our upper limits are also shown in this figure. Here we incorporate 
the completeness for the SDF sample as a function of $z^\prime$ 
magnitude (Kashikawa et al. 2004).
We expect that the number of galaxies brighter
than $z^\prime = 25.48$ ranges from $4 \times 10^{-4}$ to 0.5 and 
that brighter than $z^\prime = 26.07$ ranges from $3 \times 10^{-2}$ to 4 (figure 6). 

We may conclude that the number of UV bright 
galaxies at $z > 6.6$ is smaller than that at $z \sim 6$ even if all the star-forming 
galaxies emit the strong Lyman $\alpha$ emission. 
Bouwens et al. (2004b, 2005) surveyed for the star-forming galaxies at $z \sim$ 7--8 and 
$z \sim 10$ in the Hubble Ultra Deep Field and other deepest HST NICMOS field 
and demonstrate that the star formation rate density at $z > 6$ is decreasing with 
increasing redshift. Our result is consistent with their demonstration. 

\subsection{Further Comments}

In this paper, we used a combination of the two filters, $NB921$ and $z^\prime$.
In general, when the filter band-passes overlap, interpretations may sometimes 
get convoluted. 
However, in particular, when we study either LBGs or Lyman $\alpha$ emitters or both 
at high redshift, there is little ambiguity in selecting such galaxies as 
demonstrated in figure 4. 

Our study as well as previous works (e.g., Bouwens et al. 2005, Shimasaku et al. 2005) 
have suggested possible evidence for the decreasing
star formation activity with increasing redshift. 
In future, it seems desirable to obtain near infrared imaging data of the SDF. In particular,
if we would have a deep imaging data with a filter that starts at $\lambda =1008$ nm,
we could make a comparative study of the star formation activity between $z \sim$ 6.6 
and $z \sim 7.5$ unambiguously. The NICMOS camera on board the HST is the most efficient camera
for this kind of works. However, its field of view is too small to cover the entire
filed of the SDF. The new near infrared camera and spectrograph, MOIRCS 
(Multi-Object Infrared Camera and Spectrograph; Tokoku et al. 2003),
will be available on the Subaru Telescope; its FOV is $4^\prime \times 7^\prime$.
This camera will be useful in exploring the star formation activity at $z > 7$
in the SDF.

\vspace{1pc}
We would like to thank the staff of the Subaru Telescope. 
We would also like to thank the anonymous referee for the useful
comments and suggestions.
This work was financially supported in part by
the Ministry of Education, Culture, Sports, Science, and Technology
(Nos. 10044052, and 10304013) and JSPS (No. 15340059 and 17253001).
MA, SSS, and TN are JSPS fellows.


\clearpage

\begin{figure}
\begin{center}
\FigureFile(80mm,80mm){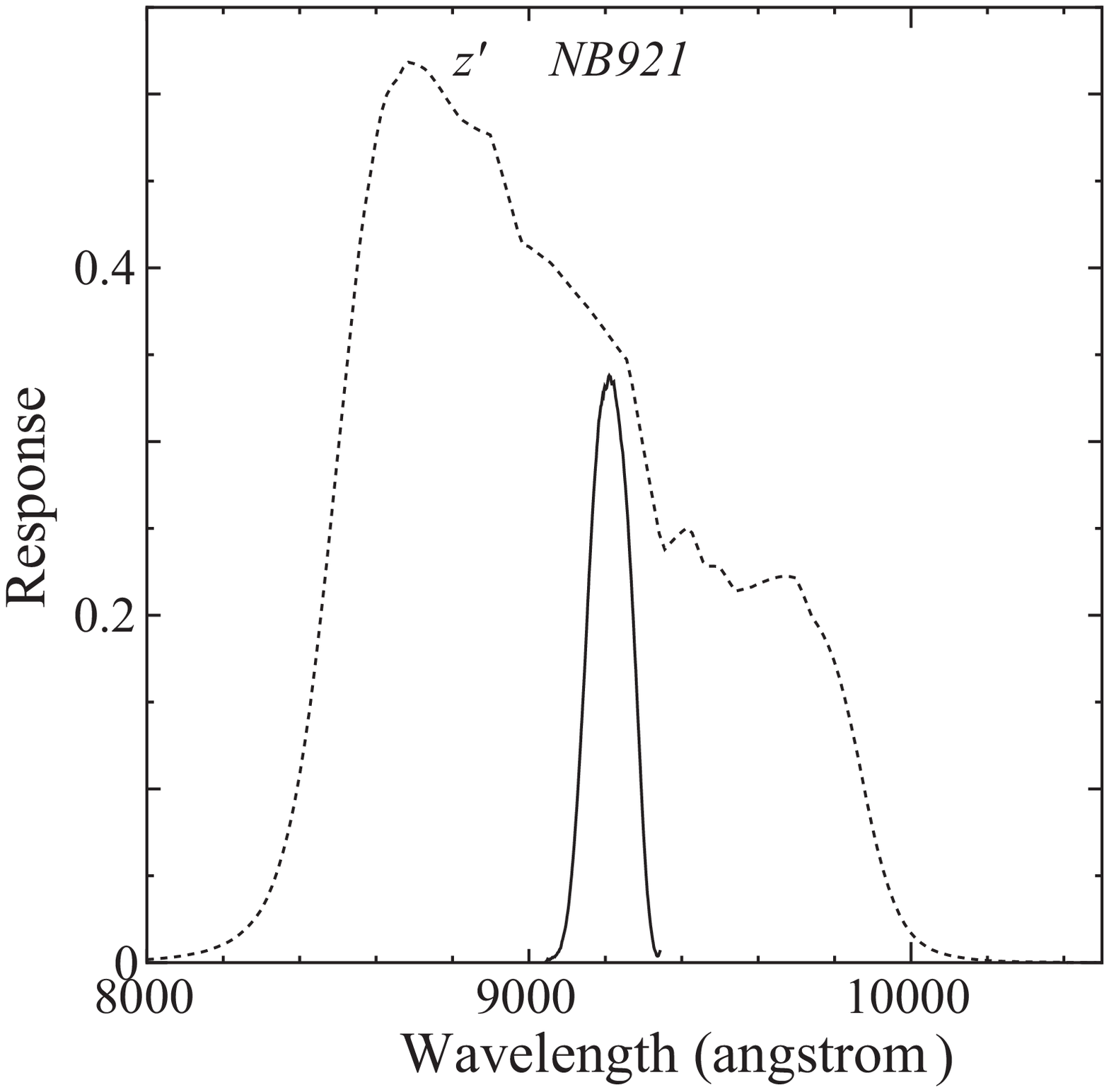}

\FigureFile(80mm,80mm){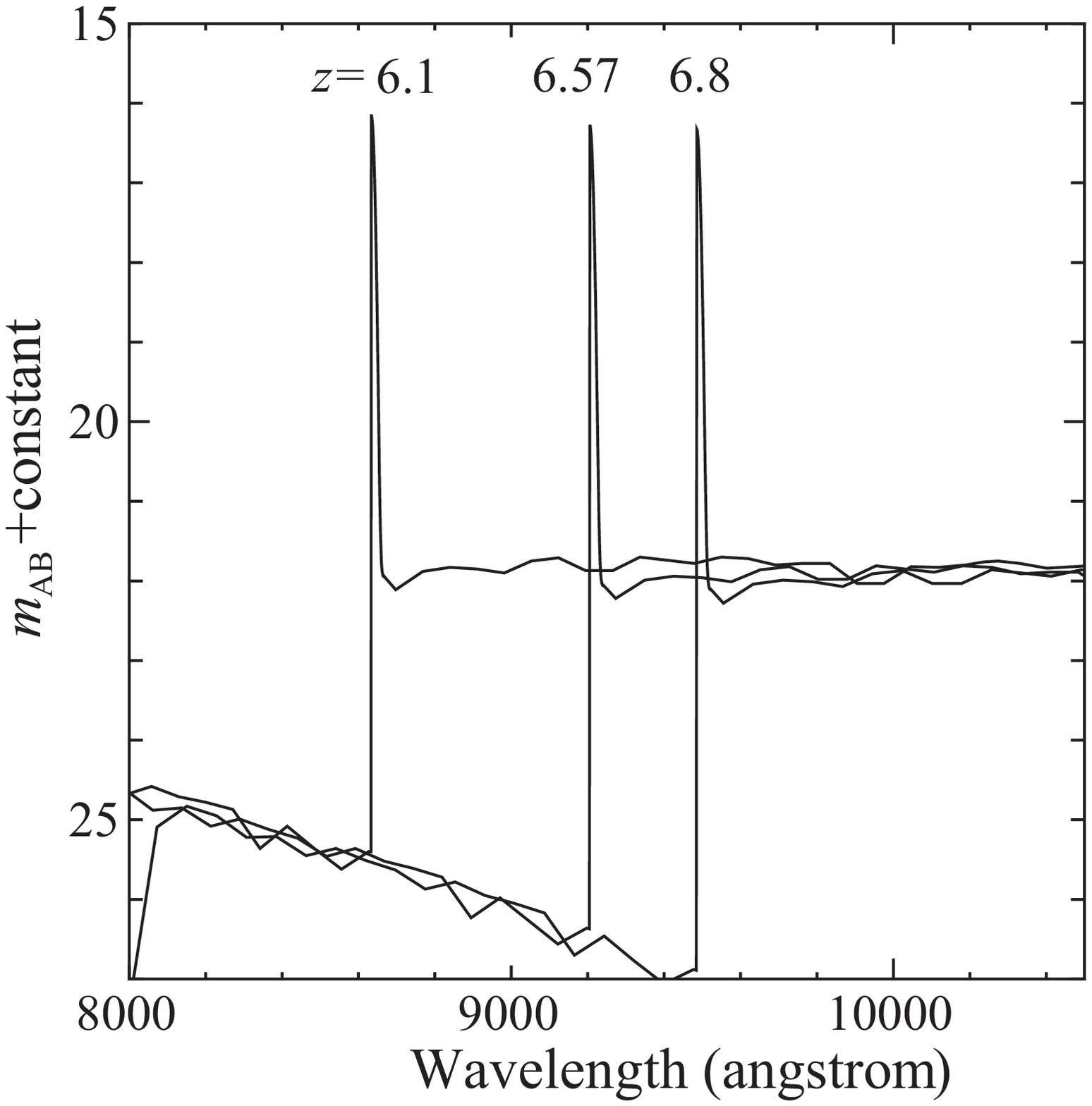}
\end{center}
\caption{
The upper panel shows the band response curves of the filters, 
$z^\prime$ and NB921. 
These are convolved with the CCD sensitivities, instrument and atmospheric transmission. 
The lower panel shows the SED of 
Lyman break galaxies at $z \sim 6.1$, 6.57, and 6.8.
}\label{fig:fig1}
\end{figure}

\begin{figure}
\begin{center}
\FigureFile(80mm,80mm){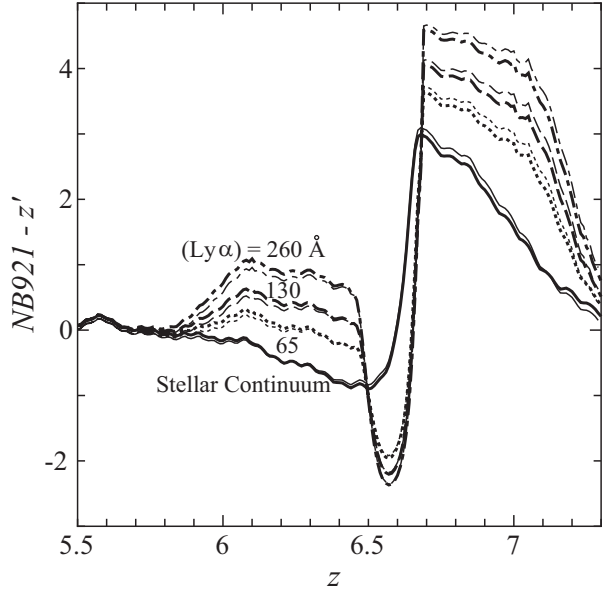}
\end{center}
\caption{
The $NB921-z^\prime$ colors for galaxies with $EW_0({\rm Ly}\alpha) =$ 
0 (solid), 65 (dotted), 130 (dashed), and 260 (dot-dashed) \AA 
as a function of the redshift. Thick (thin) lines show the case of 
$E(B-V)=0 \; (0.3)$.
}\label{fig:fig2}
\end{figure}

\begin{figure}
\begin{center}
\FigureFile(80mm,80mm){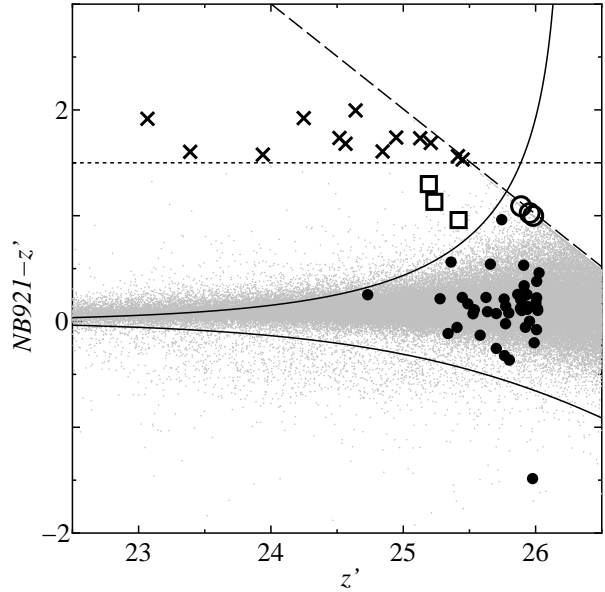}
\end{center}
\caption{
Color-magnitude diagram between $NB921-z^\prime$ and $z^\prime$. 
The horizontal dotted line corresponds the color of $NB921-z^\prime = 1.5$. 
The solid curves show the $3 \sigma$ error of $NB921-z^\prime$. 
The dashed line shows the $2 \sigma$ limiting magnitude of $NB921$. 
The $NB921-z^\prime$ colors of the objects on the line are lower limit. 
Crosses show  galaxies with $NB921-z^\prime > 1.5$. 
All of them are low redshift objects because they are detected in $B$, $V$, and $R$. 
Filled circles, open squares and open circles show 
$i^\prime$-dropout galaxies. 
Open circles show the galaxies those are not detected in NB921-band ($< 2 \sigma$). 
Open squares show the galaxies whose $NB921-z^\prime$ colors are larger than 
$3 \sigma (NB921-z^\prime)$ but smaller than 1.5. 
}\label{fig:fig3}
\end{figure}

\begin{figure}
\begin{center}
\FigureFile(80mm,80mm){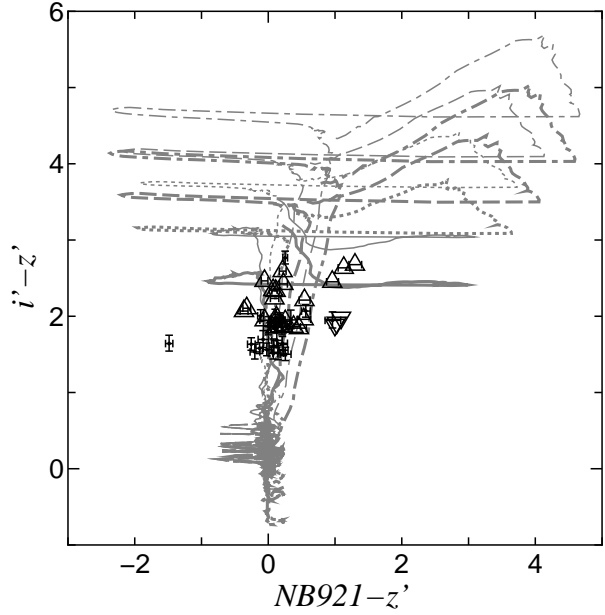}
\end{center}
\caption{
Color-color diagram between $i^\prime - z^\prime$ and $NB921-z^\prime$. 
The meanings of lines are the same as these given in Fig.2. 
All the $i^\prime$-dropouts are plotted. 
Filled circles with the error bars show galaxies detected in both $i^\prime$ and NB921 ($> 2 \sigma$), 
triangles show galaxies those are not detected in $i^\prime$-band ($< 2 \sigma$), 
and inverted triangles show those are not detected both in $i^\prime$ and NB921 ($< 2 \sigma$). 
}\label{fig:fig4}
\end{figure}

\begin{figure}
\begin{center}
\FigureFile(80mm,80mm){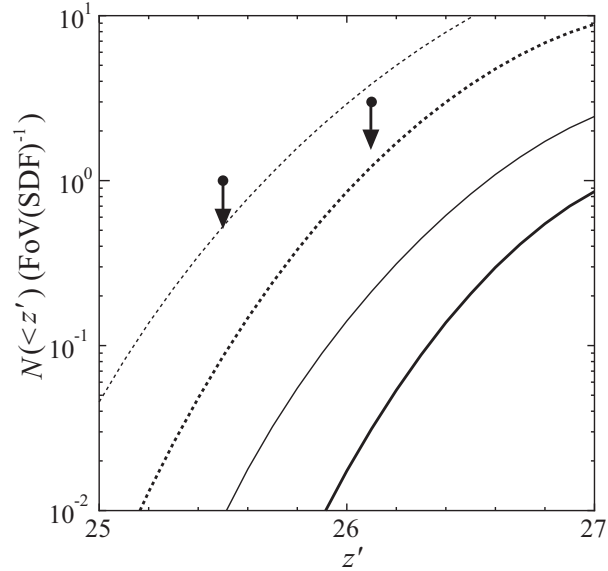}
\end{center}
\caption{
Cumulative number of NB921-dropout galaxy candidates brighter than $z^\prime = 25.48$ and 26.07. 
Lines show the the expected number of galaxies with $z > 6.6$ for different assumptions. 
Thin dotted (solid) lines show the case for the SED with (without) Ly$\alpha$ emission 
together with the UV luminosity function at $z \sim 4.0$ (Ouchi et al. 2004). 
Thick dotted (solid) lines show the case for the SED with (without) Ly$\alpha$ emission 
together with the UV luminosity function at $z \sim 6.0$ (Bouwense et al. 2004). 
}\label{fig:fig5}
\end{figure}

\end{document}